**Title:** An operational method and device for the determination of an output signal in a selected spatial section of an information processing system.

**Author:** Dr. Oliver Zafiris, study of physics and medicine at the University of Düsseldorf Germany, current address: University of Duisburg-Essen, Department for Psychiatry and Psychotherapy, Virchowstr. 174, 45147 Essen, Germany.

## Background

Current strategies in system science with a focus on neuroscience do differ in their methodological approach when exploring and trying to analyze a system in order to detect supposed underlying principle processes in its inherent actions, which one might would call 'rules' or 'laws'.

The here suggested procedure and measuring device, performs a mapping of characteristic parameters of the regional output signal, of the supposed structural properties, onto a selected regional part of the information processing system, in which the output signal and its characteristics occur.

Explicitly it is pointed out here: Here are not considered input signals, which for instance might have an influence upon (few) nuclear kernels of the atom, electrons, protons, spins of these atomic structures or substructures, or phonons, or which in general represent the physical basis for example of NMR-Physics (NMR = nuclear magnetic resonance) or solid state physics. Examples for the type of input signals considered here are visual, olfactory, tactile or auditory input signals or simply a verbal instruction.

## Concept

After the exposure to the influence of at least one complex valued input signal X during it's time interval T of influence, the suggested measuring method and procedure of data processing comprises the following units (Figure 1):

a) The measuring process of the influence of a complex valued input signal X or multiple input signals upon the information processing system,

b) the detection of at least one real valued output signal Y allocatable to a selected spatial region part of an information processing system, where the output signal Y is the result of a reaction due to the influence of an input signal X or a multitude of input signals, as described in a),

c) the creation of a family of 3-tuples of $(X, Y, Z_t)$ from the input signal X and output signals Y for the selected spatial region, where every given time interval of influence $Z_t$ is part of the time interval T of the influence (preferably precisely), a subset of IR. A 2-tupel $(X, Z_t)$ is allocatable preferably to one and only one output signal Y for the selected region part of the information processing system and (preferably exactly) one input signal uniquely allocatable to X, and

d) the determination of the reaction signal in the sense of an output signal $Y_{max}$ for the selected region of the information processing system of interest, where the reaction signal $Y_{max}$ represents the maximum or minimum of the output signals Y. Alternatively $Y_{max}$ might be a value within the interval around the maximum or minimum of the output signals Y plus or minus a positive real valued $\epsilon$, which are generated by the same input signal X within the time interval of influence T.

The time interval T is a subset of the set of real values IR, and describes the time domain of influence of the input signal X of interest, and is not predetermined and might preferable comprise the duration of onset of X. Especially it can not only span from the onset of the reaction signal till its end, but any arbitrary following temporal interval thereafter, during which the input signal is supposed of having still an influence upon the information processing system of interest.

The values of the maximum and the minimum of the output signals Y are selected as reaction signals $Y_{max}$ in two possible ways: Within an interval of values, where |.| denotes the mathematical absolute value, $Y_{max}$ is exactly the maximum of $|Y-Y_{max}|$ within the time interval T, if analytically and technically evaluable, and is selected in two possible ways: 1) The chosen maximum or minimum therefore is not placed in an interval of values $]Y_{max} - |\epsilon|, Y_{max} + |\epsilon|[$, where $\epsilon$ is a real valued number, but Y is exactly the maximum. 2) E. g. due to technical limitations of accuracy in measurement or because of analytically limitations however $Y_{max}$ might also be taken right from the open interval within $]Y_{max} - |\epsilon|, Y_{max} + |\epsilon|[$ (see Fig. 1.9).

These output values Y then may be projected onto the X-Y-plane and approximated via $f_1$: C x IR -> IR, $f=Y=Y(X, Z_t)$, where C is the set of complex numbers, by an exponential function as follows (s. Fig. 2):

$$f_1(\omega) = A - B * \exp[-\omega/\omega_a + \omega^2/\omega_{aa}], \quad (I)$$

where $\omega$ is the frequency with respect to a property of the stimulating input signals and $\omega_a$ and $\omega_{aa}$

are complex valued numbers, A, B are real valued, or for instance real valued functions of $Z_t$ only. So setting X equal to ω - for instance ω could represent the spatial frequency of a stimulus (e. g. a checker board), or ω divided by 2π, would be the value of the interstimulus interval between two following temporally sequentially input signals X which might be substructured in a complex way, and exert their influence upon the information processing system. Another example would be to define ω as the difference of contrasts of two sequentially occurring input signals divided by 2π.

For example assuming $|ω_{aa}| \gg 1$, or $|ω_a| \gg 1$, („»" denotes 'far larger than'), then the mathematical term in formula (I) can be developed according to a Taylor expansion and for example considered up to the linear or second order of ω according the selection or appropriate values for $ω_a$ or $ω_{aa}$.

A further adequate expression for the computation of appropriate term of approximating values for single 3-tuples would be $f_2$: C x IR -> IR (s. Fig. 3):

$$f_2(ω) = A - B*\exp[\sum_h^N (\prod_i (ω - ω_{0i}^h)^{p_i^h} / \prod_k (ω - ω_{0k}^h)^{q_k^h})] , \quad (II)$$

where h, i, k, N are natural numbers, $ω_{0i}^h$, $ω_{0k}^h$, are complex valued, and $p_i^h$ and $q_k^h$ are real valued, A, B defined as for formula (I).

The 3-tuples of values (X, Y, $Z_t$) are arranged as a 'spectrum', which is graphically depicted in Fig. 1.8 in a three-dimensional anticipated data curve. The resulting values for $Y_{max}$ are not necessarily positioned in a plane parallel to the X-Y-plane. A possible progression of the curve of values for Y or $Y_{max}$ into maxima, minima, saddle points or inflexion points can then represented as two-dimensional mappings (projections) and reveals locations of resonances for the input signals and their corresponding reaction signals $Y_{max}$.

Importantly one can now generate mappings of color coded statistical parameters, which base upon differences of the progression of $Y_{max}$ in the three-dimensional space or its two-dimensional projection between different regions of the same information processing system. Or one might consider differences between different information processing systems, comparing analogue spatial regions. The same is possible for the statistical comparison of values derived right from Y or $Y_{max}$ (see for instance Fig. 3.5).

If the information processing system does consist of a network of an even not manageable number of subcomponents of different materials and is not transparent in its detailed interactions then still a multitude of subcomponents might have an influence upon the regional subsection of interest generating the output signal Y. Taking into consideration the multitude of influencing structures one might consider X as system-internal or external values modified by functions like for instance the logarithm, exponential function, or a polynomial function in an appropriate order and sequence ($X^*=\log(X)$, X then is formally redefined via $X=X^*$, or the suggested procedure is performed for $X^*$, instead for X).

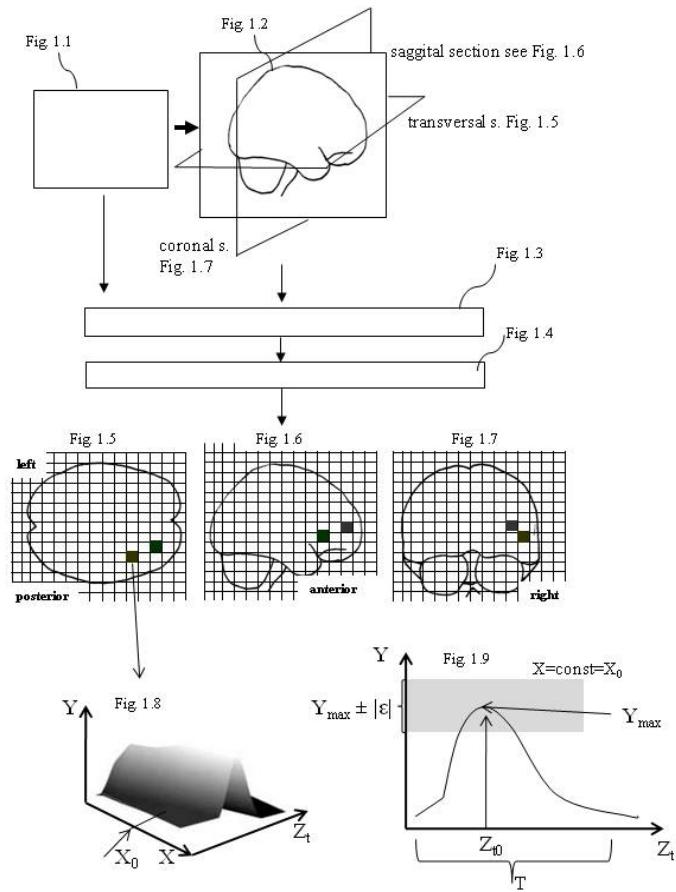

Fig. 1 depicts the measuring device which comprises at least one means for the generation and transfer of the input signals – stimuli - X (Fig. 1.1), the information processing system (Fig. 1.2, e.g. the brain), and a measurement device or means for the record of the input signals X and output signals Y (Fig. 1.3) of the information processing system. The generation of the output signals might be performed via functional magnetic resonance imaging (fMRI), magnetoencephalography (MEG), electroenecphalography (EEG), or other neuroimaging devices. The recorded signals then can be transformed into digital measurement and processing units (Fig. 1.4), and are not only visualized as centers of activation by means of color coded statistical parameters (P-, t-, F-value, Fig. 1.5, 1.6, 1.7). Importantly the regional output signals Y – here depicted by an anticipated three-dimensional data curve (Fig. 1.8) - are processed according their resulting values $Y_{max}$ and their approximation with formula (I) or (II). The resulting parameters $\omega_a$, $\omega_{aa}$, or in general e. g. $\omega_{0i}$ or $\omega_{0k}$, are then color coded and mapped onto the corresponding spatial unit.

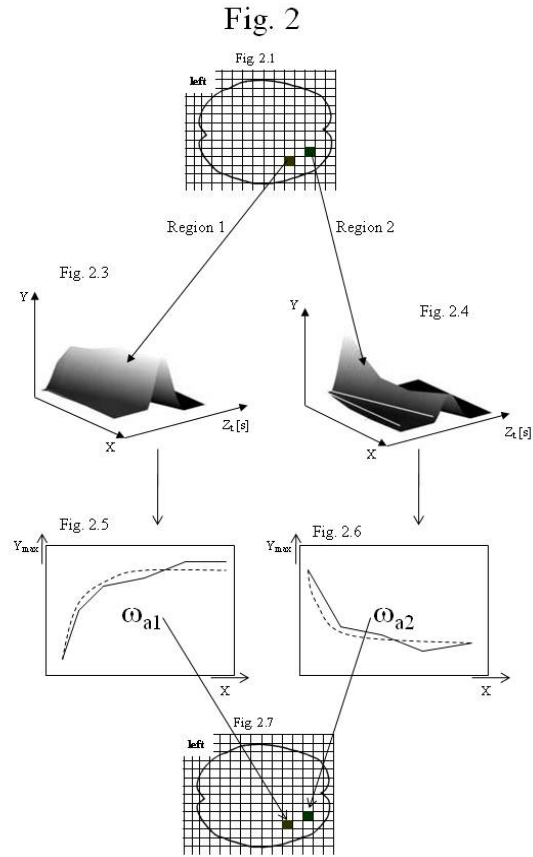

Legend to Figure 2

Two different three-dimensional anticipated data curves for two different regions of the information processing system (motivated by Science, Vol. 287, pp. 1269-72, 2000), can differ in their corresponding data curves for $Y_{max}$, although their statistical significance of activation might be (nearly) equal. The data curves of $Y_{max}$ of these regions perhaps are not necessarily in parallel to the X-Y-plane. In the same way they might still do then differ in their projections as two-dimensional data curves onto the X-$Z_t$-, X-Y- or Y-$Z_t$-plane (Fig. 2.5, 2.6). The approximation of these curves for example with formula (I) then yields different values $\omega_{a1}$ (Fig. 2.5) and $\omega_{a2}$ (Fig. 2.6), which are then mapped color coded onto the corresponding brain regions (Fig. 2.7).

Fig. 3

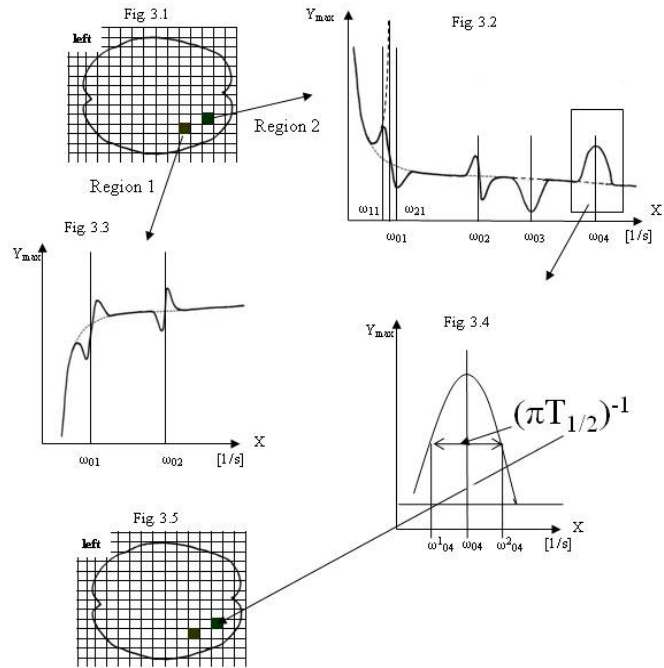

*Legend to Figure 3*

Figure 3.1 depicts two different regions of the information processing system, which differ concerning their simulated data curves for the values of $Y_{max}$ after projection onto the X-Y-plane (Fig. 3.2, 3.3). These data curves then can be approximated by formula (I) or (II). Regional signal enhancement or decrease, as exemplified in Figure 3.4, then can be processed in order to evaluate further parameters to characterize the region of interest of the information processing system (e.g. brain). The regional maximum of the approximated function $Y_{max} = f(\omega, Z_t)$ at the location $X = \omega_{04}$ suggests the definition of the value for the half-width (full width at half maximum) $T_{1/2}$ according: $\omega^2_{04}(Y=Y_{max}/2)$ minus $\omega^1_{04}(Y=Y_{max}/2)$ equal to $(\pi T_{1/2})^{-1}$ (Fig. 3.4). The values $T_{1/2}$ then can be color coded and mapped onto the corresponding activated region of the information processing system (Fig. 3.5). In the same way, values $\omega_{01}, \omega_{02, ...,}$ motivated in an anticipated data curve (Fig. 3.2, 3.3) can be mapped onto the region of the information processing system.